\documentclass[%
 amsmath,amssymb,
 aps,
prb,
floatfix,
twocolumn
]{revtex4-1}

\usepackage{CJKutf8}
\usepackage{array}
\newcolumntype{C}[1]{>{\centering\let\newline\\\arraybackslash\hspace{0pt}}m{#1}}

\usepackage{graphicx}% Include figure files
\usepackage{color}
\usepackage{dcolumn}% Align table columns on decimal point
\usepackage{bm}% bold math
\usepackage{wasysym}
\usepackage{bbold}
\usepackage{hyperref}% add hypertext capabilities

\begin{document}

\title{Spin polarization control in a 2-dimensional semiconductor}

\author{Ian Appelbaum}
\email{appelbaum@physics.umd.edu}
\author{Pengke Li (\begin{CJK*}{UTF8}{gbsn}李鹏科\end{CJK*})}
\email{pengke@umd.edu}
\affiliation{Department of Physics and Center for Nanophysics and Advanced Materials, U. of Maryland, College Park, MD 20742}

\begin{abstract}
Long carrier spin lifetimes are a double-edged sword for the prospect of constructing ``spintronic" logic devices: Preservation of the logic variable within the transport channel or interconnect is essential to successful completion of the logic operation, but any spins remaining past this event will pollute the environment for subsequent clock cycles. Electric fields can be used to manipulate these spins on a fast timescale by careful interplay of spin-orbit effects, but efficient controlled depolarization can only be completely achieved with amenable materials properties. Taking III-VI monochalcogenide monolayers as an example 2D semiconductor, we use symmetry analysis, perturbation theory, and ensemble calculation to show how this longstanding problem can be solved by suitable manipulation of conduction electrons.
\end{abstract}

\maketitle

\section{introduction}
Manipulation of electron spin orientation in polarized ensembles provides a basis for new logic devices and circuits with potential advantages over present-day charge-based designs.\cite{Dery_HSTM12} It is widely believed that whenever spin encodes logic state, semiconductor materials with the longest spin lifetime are the most suitable choice for transport channels between injection and detection contacts. However, once a logic operation is completed, residual spins can -- and will -- interfere with those involved in future operations. Can we design a device with a controllable spin lifetime? In this scheme, otherwise robust spins would vanish from the channel by an externally-induced, fast, and tunable depolarization mechanism upon completion of every logic operation.

The present paper presents a solution to this challenge, making use of two-dimensional semiconductor materials having a strong \emph{uniaxial} spin-orbit field anisotropy. In this scheme, spins are initially aligned parallel or antiparallel to a long-lived quantization axis at injection or generation. After spin transport to other parts of the device and completion of a logic operation, a clocked voltage pulse at an electrostatic gate generates an electric field in the transport channel that induces a Bychkov-Rashba effective magnetic field.\cite{Bychkov_JETPL84} This magnetic field, due to the structural inversion symmetry-breaking electric field and spin-orbit interaction, is non-collinear to the spin axis and thus rotates the spins via precession onto an orthogonal axis.  The physical logic environment is then reset for the next operation.

One realization of such an anisotropic material is the zincblende [110] quantum well, whose spin relaxation properties have been thoroughly studied using optical orientation methods.\cite{Ohno_PRL99, Dohrmann_PRL04, Couto_PRL07, Muller_PRL08} However, fabrication of this system requires epitaxial growth and the active layer is buried deep within the bulk. An alternative approach to meet our requirement for anisotropy without sophisticated crystal growth or the constraint of deep encapsulation is through use of inversion-asymmetric van~der~Waals layered materials obtainable by exfoliation or vapor deposition methods. 

Through detailed theoretical symmetry analysis, we have identified several such two-dimensional materials with the requisite anisotropic spin-orbit properties. The most promising candidate material system we have found is the group-III metal--{\em mono-}chalcogenide monolayers (G3M-MCs). In this inversion-asymmetric two-dimensional material system (such as GaSe and InS), the spin-orbit-induced $\mathbf{k}$-dependent Dresselhaus effective magnetic field\cite{Dresselhaus_PR55} is oriented perpendicular to the monolayer plane and scales as a cubic function of the wavenumber $k$.\cite{Li_PRB15} Spin up and down are then the natural eigenstates, immune to Dyakonov-Perel (DP) spin relaxation which would otherwise cause precessional dephasing upon momentum scattering for any other polarization axis.\cite{Dyakonov_SPSS72}  An electrostatically-controlled Bychkov-Rashba field -- which is always perpendicular to both the quasimomentum $\mathbf{k}$ and the electric field $\mathcal{E}\mathbf{z}$, thus oriented in-plane -- can be used to rotate spins toward the plane and achieve depolarization. 

\begin{figure}
\includegraphics{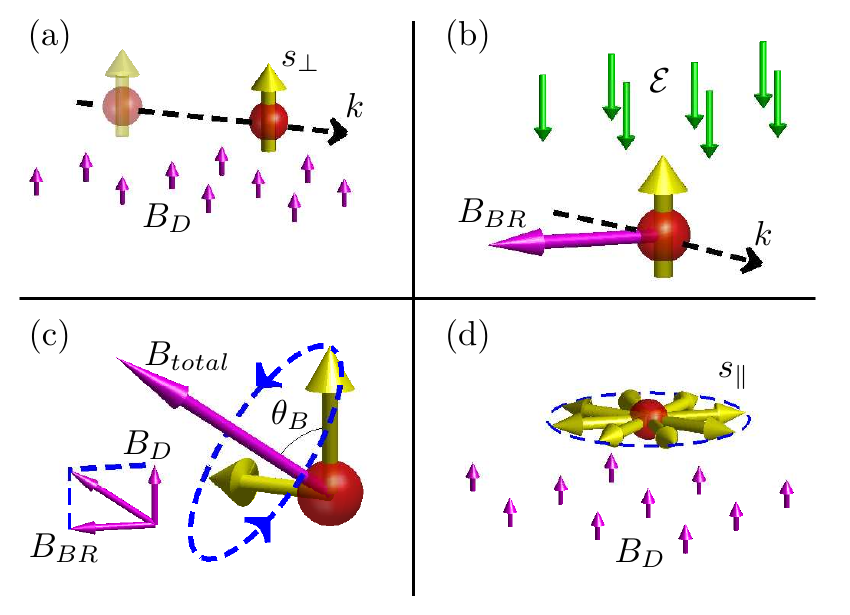}
%\vspace{-10pt}
\caption{Spin polarization control using pulsed spin-orbit fields. Panel (a) shows a charge carrier with quasimomentum $k$ and long-lived spin $s_\perp$, perpendicular to the plane and parallel to the Dresselhaus magnetic field $B_D$. In (b), electric field $\mathcal{E}\parallel s_\perp$ creates a Bychkov-Rashba spin-orbit field $B_{BR}\perp \mathcal{E}, k$ in the plane. Spins then precess about the total field $B_{total}$ at angle $\theta_B$, as shown in (c). When the electric field vanishes, any residual in-plane spins $s_\parallel$ are quickly dephased by the Dresselhaus field, as shown in (d).   \label{fig:scheme} }
\end{figure}

\section{Mechanism and materials}

The depolarization mechanism we describe is illustrated in Fig. \ref{fig:scheme}(a)-(d). In (a), spin-polarized electrons oriented normal to the channel surface are injected electrically from a ferromagnet with perpendicular magnetic anisotropy (such as the CoFeB/MgO system\cite{Ikeda_NatMater10} and Co/Ni or Co/Pd ultrathin multilayer system\cite{Carcia_JAP88, *Daalderop_PRL92}) or generated via optical interband excitation with polarized light.\cite{Li_PRB15} The out-of-plane spin-orbit Dresselhaus field stabilizes spins aligned (or anti-aligned) to it from extrinsic fluctuations (such as magnetic impurities, random strain gradient, substrate potential fluctuation, etc), allowing them to travel through the channel without appreciable depolarization. In Fig. \ref{fig:scheme}(b), we show that after a logic operation is completed (such as spin-torque or exchange from polarized electrons manipulating the magnetization of a ferromagnetic contact\cite{Yang_NatPhys08}), a perpendicular electric field pulse provided by a transverse electrostatic gate induces a Bychkov-Rashba effective magnetic field oriented in the plane. Its combination with the intrinsic Dresselhaus field results in a total effective magnetic field misaligned with the spins, at an angle $\theta_B$. In Fig. \ref{fig:scheme}(c), we show spin precession around the total spin-orbit field. With a carefully engineered gate voltage pulse amplitude and duration, spins precess into the plane, eliminating the ensemble projection onto the original quantization axis when the electric field vanishes. As shown in Fig. \ref{fig:scheme}(d), the channel is then cleared of out-of-plane spin, and any remaining in-plane polarization is quickly depolarized by precessional dephasing. Residual spins are eliminated, preparing the channel for the next logic cycle (which may be affected by the up/down orientation of the injector ferromagnet from upstream circuit elements).

Several questions must be answered before this scheme can be considered viable: Which perpendicularly-polarized carriers (conduction electrons or valence holes, immune to DP) suffer the least relaxation by secondary spin-flip mechanisms? What is the magnitude of both the Dresselhaus and Bychkov-Rashba coefficients for this band, and are they compatible to achieve complete depolarization in electric fields of reasonable strength? What is the relationship between optimized gate pulse duration and electric field, and is it consistent with the requirements imposed by an upper bound set by momentum scattering time? In the following sections, we apply symmetry analysis, lowest-order perturbation theory, and ensemble integration to address these and other essential questions.   

Before proceeding to the next section, we must first address an important issue regarding our choice to focus on the lesser-known G3M-MC monochalcogenide materials (GaSe, InS, etc.), in contrast to the better-known TMDC {\em di-}chalcogenide system (WS$_2$, MoSe$_2$, etc.). In monolayers of both materials, the internal Dresselhaus magnetic field (proportional to the spin-subband splitting) is always oriented out-of-plane.\cite{Yao_PRB08, *Xiao_PRL12} However, gap-edge states of G3M-MCs are located around the Brillouin zone center $\Gamma$-point, while those of TMDCs are at the zone-edge $K(K')$ points. As a result, the internal spin-orbit effective magnetic field dependence on crystal momentum $k$ is drastically different in each case: spin splitting scales as $k^3$ in G3M-MCs, vanishing at $\Gamma$ due to Kramers' degeneracy there, but in TMDCs, the spin splitting close to the band extrema is enormous in magnitude and independent of the wavevector $k$. The smallest spin-splitting can be found in the MoS$_2$ conduction band at 4~meV, equivalent to a magnetic field of many tens of tesla; to compete with it, the voltage-induced Bychkov-Rashba field will require similarly enormous and impractically obtainable electric fields. %Furthermore, strong $K-K'$ intervalley scattering of in-plane polarized electrons will lead to motional narrowing due to fluctuating time-reversed fields, weakening the intrinsic relaxation anisotropy.

\section{perpendicular spin lifetime}
We first justify our expectation of a long out-of-plane spin lifetime, and motivate the choice of conduction-band electron manipulation in n-type G3M-MC monolayers, as opposed to holes in p-type material. 

\subsection{Spin mixing}
Although spins aligned to the Dresselhaus field are immune to DP relaxation, they are still subject to the Elliott-Yafet (EY) mechanism. EY spin relaxation is driven by carrier scattering events that couple to minority spin components of the wavefunction. These impure admixtures are introduced by the effect of spin-orbit interaction and so EY is generally present in all materials regardless of inversion (a)symmetry.  

Spin-orbit interaction can be treated as a perturbation within $\mathbf{k}\cdot\mathbf{\hat{p}}$ theory, where it generates two terms in the envelope Hamiltonian: a $\mathbf{k}$-independent term $\frac{\hbar}{4m_0^2c^2}\nabla V\times\mathbf{\hat p}\cdot\vec{\sigma}$ and a $\mathbf{k}$-dependent term $\frac{\hbar^2}{4m_0^2c^2}\nabla V\times\mathbf{\hat k}\cdot\vec{\sigma}$. In two-dimensional systems when $\mathbf{k}\cdot\mathbf{z}=0$, the latter takes on the form 
\begin{align}
\frac{\hbar^2}{4m_0^2c^2}\left[(k_x\sigma_y-k_y\sigma_x)\frac{\partial V}{\partial z}+k_y\sigma_z\frac{\partial V}{\partial x}+k_x\sigma_z\frac{\partial V}{\partial y}\right].
\label{eq:k_dep}
\end{align}
Since only $\sigma_x$ and $\sigma_y$ have off-diagonal elements, only the first term (proportional to $\frac{\partial V}{\partial z}$) can perturb the wavefunction with opposite spin admixtures. This term clearly has the same spatial symmetry properties as the polar vector $z$; using the language and notation of group theory, it is a basis function for the irreducible representation (IR) $\Gamma_2^-$, as in Table~\ref{tab:BasisFunctions}. 

\begin{table} [h!]
\caption{Basis functions (BFs) of some irreducible representations (IRs) in $\Gamma$-point $D_{3h}$ group.  
The assignment of plus and minus superscripts to IRs follows the convention of even and odd parity with respect to the operation of in-plane mirror reflection $\sigma_h$.}
\label{tab:BasisFunctions}
\renewcommand{\arraystretch}{1.1}
\begin{tabular}{C{1.2cm}|C{1.2cm}C{1.2cm}C{1.2cm}C{1.5cm}}
\hline \hline
IRs&$\Gamma_1^+$&$\Gamma_3^+$&$\Gamma_2^-$&$\Gamma_3^-$
\\ \hline
BFs&$\mathbb{1}$&\{$x$, $y$\}&$z$&\{$xz$, $yz$\}
\\
\hline
\end{tabular}
\end{table}

All components of the operator $\mathbf{\hat p}$ exist regardless of the dimensionality, so the $\mathbf{k}$-independent spin-orbit interaction is
\begin{align}
\frac{\hbar}{4m_0^2c^2}&\left[\sigma_x\left(\frac{\partial V}{\partial y}p_z-\frac{\partial V}{\partial z}p_y\right)+\sigma_y\left(\frac{\partial V}{\partial z}p_x-\frac{\partial V}{\partial x}p_z\right) \right.\notag \\
&\left. +\sigma_z\left(\frac{\partial V}{\partial x}p_y-\frac{\partial V}{\partial y}p_x\right)\right].
\label{eq:k_indep}
\end{align}
The $\sigma_x,\sigma_y$ spin-mixing terms transform as the in-plane components of an axial (pseudo-) vector $\{ xz,yz \}$ (the IR $\Gamma_3^-$, see Table~\ref{tab:BasisFunctions}).

With the assistance of the basis functions (Table~\ref{tab:BasisFunctions}) that capture the symmetries of spin-orbit perturbations and different bands, it is straightforward to examine how spin-mixing is introduced. Here we focus on the gap-edge states. The valence band spatial wavefunctions are invariant to all of the point-group symmetry operations, and thus transform as a scalar, $\mathbb{1}$ (corresponding to IR $\Gamma_1^+$). The $\mathbf{k}$-\textit{dependent} spin-orbit perturbations  thus cause first-order corrections of opposite spin from remote bands with $\Gamma_2^-$ ($z$-like) symmetry, since $\langle \mathbb{1} | \frac{\partial V}{\partial z} |z\rangle$ is nonvanishing. Similarly, opposite spin components are induced to the valence band by $\mathbf{k}$-\textit{independent} spin-orbit perturbations from remote bands with $\Gamma_3^-$ ($\{xz,yz\}$-like) symmetry. The same argument can be applied to the conduction band, which is odd with respect to mirror inversion about the plane, and so transforms like $z$ ($\Gamma_2^-$). The conduction band wavefunction will thus acquire spin admixtures with spatial symmetries of $\Gamma_1^+$ from $\mathbf{k}$-dependent and $\Gamma_3^+$ from $\mathbf{k}$-independent perturbations. 

% The valence band spatial wavefunctions are invariant to all of the point-group symmetry operations, and thus transform as a scalar, $\mathbb{1}$ (belonging to IR $\Gamma_1^+$). The spin-orbit perturbations above thus cause first-order corrections of opposite spin having $\{xz,yz \}\otimes\mathbb{1}\rightarrow \{xz,yz \} (\Gamma_3^-)$ symmetry ($\mathbf{k}$-independent) and $z\otimes\mathbb{1}\rightarrow z (\Gamma_2^-)$ symmetry ($\mathbf{k}$-dependent). The conduction band is odd with respect to mirror inversion about the plane, and so transforms like $z$ ($\Gamma_2^-$ ). These wavefunctions will acquire spin admixtures with spatial symmetries of $z\cdot\{ xz,yz\}\rightarrow\{ x,y\}(\Gamma_3^+)$ ($\mathbf{k}$-independent) and $z\cdot z\rightarrow \mathbb{1}(\Gamma_1^+)$ ($\mathbf{k}$-dependent), where we recognize the fact that $z^2$ is invariant to all symmetry operations and thus is merely a higher-order basis function alternative to $\mathbb{1}$ for $\Gamma_1^+$. 

\subsection{Phonon symmetry}

In-plane acoustic phonons in these materials have  {$x,y$} ($\Gamma_3^+$) symmetry and therefore only play a role in spin-{\em {preserving}} momentum scattering. These events couple the spin-majority  components of the wavefunctions
%with a perturbation induced by the last two terms in the $\mathbf{k}$-dependent spin-obit interaction in Eq.~(\ref{eq:k_dep}), 
and affect the charge mobility but not spin relaxation. In the following we discuss the influence on spin relaxation due to carrier scattering with flexural phonons and optical phonons, and justify that in both cases, the spin of electrons in the conduction band is more robust than holes in the valence band.

Because out-of-plane flexural phonons have no cut-off and a quadratic dispersion relation to lowest order (and hence a constant density of states, as opposed to the vanishing linear DOS for the in-plane acoustic phonons), scattering with them usually dominates the EY spin lifetime in two-dimensional materials.\cite{Song_PRL13, Fratini_PRB13, Han_NNano14} These phonons have spatial symmetry of $z$ ($\Gamma_2^-$) and so will drive spin relaxation in both the valence and conduction bands by coupling majority spin to admixtures introduced by the spin-flip terms of the $\mathbf{k}$-dependent spin-orbit interaction in Eq.~(\ref{eq:k_dep}). 

The conduction band dispersion is quadratic around the Brillouin zone center, so thermal electrons filling these states will have very small $k$ and thus negligible $\Gamma_1^+$ spin admixtures. The valence band, however, has a distorted `caldera' shape and so thermally occupied holes at the bandedge -- on the caldera rim -- have a substantial $k$. This in turn leads to a strong wavefunction admixture with components having $\Gamma_2^-$ character. As a result, we expect that the valence band states will be far more susceptible than conduction band states to spin relaxation caused by the unavoidable presence of flexural phonons. 

The $\mathbf{k}$-independent perturbation Eq.~(\ref{eq:k_indep}) exacerbates the problem for holes. This spin-orbit term leads to spin flips in both conduction and valence bands via scattering with in-plane optical phonons sharing the same $\Gamma_3^-$ symmetry of the two spin-mixing terms in Eq.~(\ref{eq:k_indep}). The cutoff energy of this type of phonon in G3M-MC monolayers is $\approx$25 meV \cite{Bashenov_PSS78, *Altshul_PSS80} and is therefore expected to seriously affect spin relaxation at room temperature. The strength of EY spin relaxation due to scattering with these optical phonons is proportional to the minority-spin mixing amplitude of the eigenstates, which is far larger in the $\Gamma_1^+$ highest valence band (8\% probability as opposed to 0.1\% in the conduction band)\cite{Li_PRB15} due to the close proximity of $\Gamma_3^-$ lower valence bands.

In light of these issues, we conclude that the spin lifetime for valence band holes is much shorter than electrons in the conduction band of G3M-MC monolayers. We therefore restrict our subsequent analysis to the latter carriers. For electrical injection of spin-polarized electrons into the conduction band, n-type conductivity is desirable, as is usually the case in GaS\cite{Late_Adv_Mat12} and InSe\cite{Mudd_Adv_Mat15, *Sucharitakul_Nano_Lett15}, whereas GaSe is usually p-type.\cite{Late_Adv_Mat12,Li_SciRep14} 
Controllable n-doping during synthesis is therefore desirable in this case. On the other hand, spin injection via optical orientation during photocarrier generation is insensitive to the doping nature, while electron spin relaxation due to exchange with holes (Bir-Aronov-Pikus mechanism\cite{BAP_JETP75}) should play a minor role thanks to the relatively spin-pure gap edge states (as compared with degenerate valence edge states in cubic systems). In both doping cases, back-gate bias tuning may be necessary to reduce the background Bychkov-Rashba field induced from structural inversion asymmetry introduced by the presence of the substrate.

\section{Conduction band Dresselhaus and Bychkov-Rashba coefficients}
The proposed mechanism to exploit the spin-orbit anisotropy for channel reset depends crucially on our ability to generate an in-plane Bychkov-Rashba field that rivals the out-of-plane Dresselhaus field in magnitude. Only then will a sufficient component of spin precess into the orthogonal in-plane orientation. Here, using third-order perturbation theory to calculate the magnitudes of these two fields, we demonstrate the feasibility of this scheme. 

\begin{figure}
\includegraphics[width=7cm]{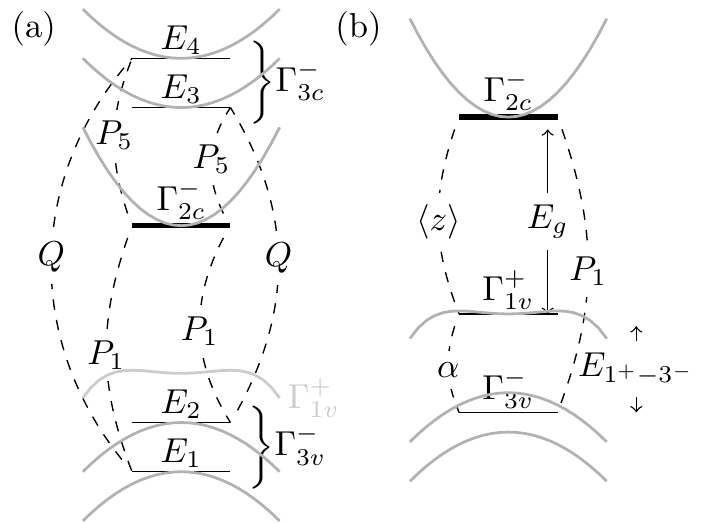}
\caption{Matrix-element perturbation pathways allowing calculation of spin splitting in the $\Gamma_2^-$ conduction band of monochalcogenide monolayers via two distinct mechanisms: (a) ``bulk inversion asymmetry" Dresselhaus coefficient, and (b) ``structural inversion asymmetry" Rashba coefficient. Not to scale. \label{fig:Dresselhaus_Rashba}}
\end{figure}

First of all, following the same scheme of evaluating the valence band Dresselhaus spin splitting,\cite{Li_PRB15} we can estimate the magnitude $\gamma_c$ of the conduction band Dresselhaus term $\mathcal{H}_D=\gamma_c k^3 \sin 3\phi \sigma_z$ using perturbation theory to third order in $\mathbf{k\cdot\hat{p}}$. The dominant terms, reminiscent of those in the analogous calculation for III-V semiconductors \cite{Cardona_PRB88}, correspond to perturbation paths through both the spin-orbit-split lower valence and upper conduction bands. Referring to Fig. \ref{fig:Dresselhaus_Rashba}(a), where horizontal lines represent the spin-dependent $\Gamma$-point states, one obtains a magnitude
\begin{align}
\gamma_c&=\frac{\hbar^3}{m_0^3}\sum_{i,j = \Gamma_{3\!v,c}^-}\frac{\langle\Gamma_{2c}^-|p_y|i\rangle\langle i|p_y| j\rangle\langle j| p_y|\Gamma_{2c}^-\rangle}{E_iE_j}\nonumber\\
&=|P_1QP_5|\left(\frac{1}{E_2E_3}-\frac{1}{E_1E_4}\right),
\label{eq:gamma1}
\end{align}
where $P_{1}$ and $P_{5}$ are proportional to off-diagonal matrix elements of the momentum operator $\hat{p}_{x,y}$, i.e. $P_{1(5)} = \frac{\hbar}{m_0}\langle\Gamma_{2c}^-|\hat{p}_{x,y} |\Gamma_{3v(c)}^-\rangle$, and $Q$ is the matrix element $\frac{\hbar}{m_0}\langle\Gamma_{3v}^-|\hat{p}_{x,y} |\Gamma_{3c}^-\rangle$. $E_1$ and $E_2$ ($E_3$ and $E_4$) are the energies of the spin-split $\Gamma_{3v}^-$ ($\Gamma_{3c}^-$) bands relative to $\Gamma_{2c}$.

Calculation of the electrostatic gate-induced Bychkov-Rashba coefficient can be treated similarly within perturbation theory. As shown in Fig. \ref{fig:Dresselhaus_Rashba}(B), the dominant path is via the closest $\Gamma_{1v}^+$ and $\Gamma_{3v}^-$ valence bands. Here, the out of plane electric field $\mathcal{E}\mathbf{z}$ directly couples the gap edge $\Gamma_{2c}^-$ and $\Gamma_{1v}^+$ states because they are of opposite reflection parity. The $\mathbf{k}\cdot\hat{\mathbf{p}}$ perturbation [with the same parameter $P_1$ as in Eq.~(\ref{eq:gamma1})] strongly couples the $\Gamma_{2c}^-$ and $\Gamma_{3v}^-$ states that share the same in-plane planewave origin.\cite{Li_PRB15} Coupling between the two intermediate states is by the $\mathbf{k}$-independent spin-orbit term  in Eq. \ref{eq:k_indep}, which is related to the strong spin-mixing coefficient $\alpha_v$ of the $\Gamma_{1v}^+$ valence band. The Bychkov-Rashba coefficient can then be evaluated by
\begin{align}
\beta_c\approx|P_1e\mathcal{E}\langle z\rangle\alpha_v|\frac{E_{1^+-3^-}}{E_g^2},
\label{eq:beta1}
\end{align}
where $\langle z\rangle$ is on the order of the monolayer thickness and $E_{1^+-3^-}$ is the energy difference between $\Gamma_{1v}^+$ and $\Gamma_{3v}^-$. Depending on the average wavevector of the electrons, the Bychkov-Rashba term $\mathcal{H}_{BR}=\beta_c k(\cos\phi\sigma_y-\sin\phi\sigma_x)$ can be tuned from zero up to a value comparable with (or even dominant over) the weak Dresselhaus term $\gamma_c k^3\sin 3\phi\sigma_z$. %It is essential here that the symmetry of G3M-MC monolayers strictly precludes the existence of Dresselhaus terms linear in $k$, otherwise the required electric field for device operation would be unreasonably large; see next section on ensemble calculations.   

Using parameters appropriate for the conduction band of monolayer GaSe ($\alpha_v\approx 0.3$, $E_g\approx3$~eV, $P_1\approx h^2/ma$, $a\approx 3.75$\AA,\, $E_{1^+-3^-}\approx0.3$~eV), we obtain an expected Bychkov-Rashba energy on the order of 1~meV for electrons with $k=\sqrt{2m^*k_BT}/\hbar\approx 0.1\frac{\pi}{a}$ at $T=300$~K in an electric field $\mathcal{E}=1$~MV/cm, readily obtainable with thin-film dielectric insulators and low gate voltages\cite{Gupta_EDL97}, and of the same order as the Dresselhaus splitting at the same $k$ along $\Gamma-K$ calculated from a tight-binding bandstructure calculation
(cubic polynomial fitting the Dresselhaus spin-split dispersion gives $\gamma_c\approx$1.044~eV\AA$^3$).\cite{Li_PRB15, Robertson_JPC79,Camara_PRB02,Chadi_PRB77} By incorporating on-site electrostatic energy into the calculation, we can fit the linear conduction band splitting along $\Gamma-M$ (where Dresselhaus effect vanishes) to recover $\beta_c$: For Ga(Se) atoms 1.2(2.3)\AA\,\cite{Rybkovskiy_PRB14} from the basal plane, our numerical results yield $\beta_c/\mathcal{E}\approx$2.4~meV\AA/(MV/cm).

\section{ensemble summation}

Of course, not all electrons have the same $\mathbf{k}$ and hence feel different Dresselhaus and Bychkov-Rashba fields. Thus, the shortest possible electrostatic gate pulse-width optimizing precession-induced depolarization of the out-of-plane component of these electrons is dependent on which states comprise the nondegenerate electron density $n=D_{2d}k_BT \exp(E_F/k_BT)$, where $D_{2d}$ is the (constant) density of states, $k_BT$ is the thermal energy, and $E_F<-k_BT$ is the Fermi energy relative to the conduction band minimum.

An initially out-of-plane spin precesses around an effective magnetic field at an angle $\theta_B$ with the surface normal, and therefore has an out-of-plane projection
\begin{align}
S_z(\omega t,\theta_B)&=\cos\omega t\sin^2 \theta_B + \cos^2\theta_B.\label{eq:sz}
\end{align}
The in-plane spin components are 
\begin{align}
S_x(\omega t,\theta_B)&=\sin\omega t\sin \theta_B,\notag\\
S_y(\omega t,\theta_B)&=\sin^2\frac{\omega t}{2}\sin 2\theta_B,\notag
\end{align}
where the $y$-direction lies in the plane formed by the initial spin vector and the effective magnetic field.

\begin{figure}
\includegraphics[width=7.5cm]{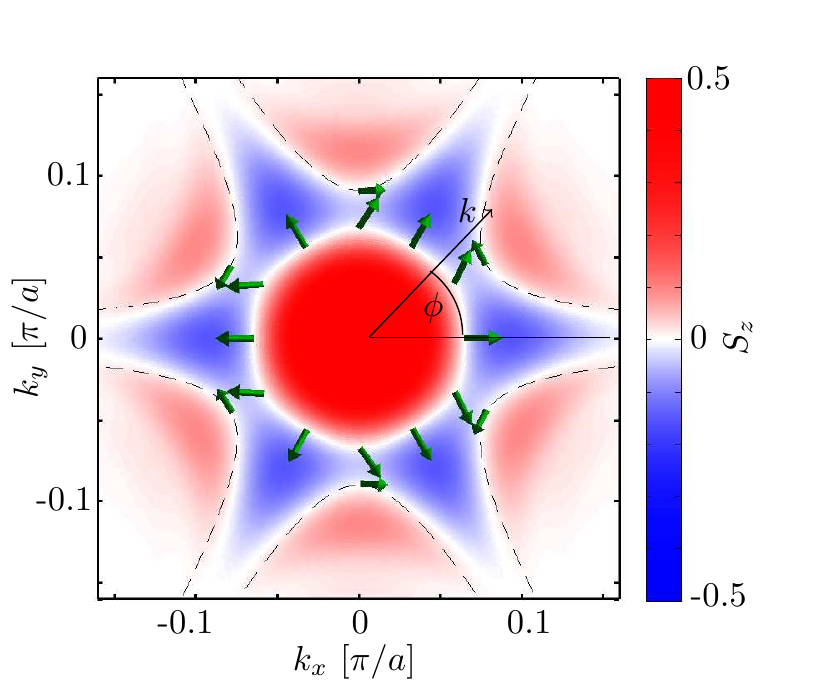}
\caption{Spin orientation in $k$-space for a thermal ensemble of initially perpendicularly-polarized spins at $T=300$~K, with $\beta_c=$6~meV\AA, at $t=\pi\hbar/\sqrt{2\beta_c^3/\gamma_c}\simeq 3.22$~ps. In-plane spin vectors are plotted for several states where $S_z=0$. Dashed curves mark $k=\sqrt{\beta_c/(\gamma_c\sin{3\phi})}$, where $\theta_B=\pi/4$. Here, $\gamma_c=1.044$~eV\AA$^3$  and $m^*=0.655m_0$, as appropriate for the conduction band of monolayer GaSe.
\label{fig:BZ}
}
\end{figure}

Because the spin-orbit Hamiltonian terms $\mathcal{H}_{BR}(k,\phi)$ and $\mathcal{H}_D(k,\phi)$ are time-reversal-invariant, equilibrium ensemble averages over the in-plane components $\langle S_x\rangle$ and $\langle S_y\rangle$ for initially perpendicularly-polarized spins vanish identically for all $t$. This can clearly be seen in Fig. \ref{fig:BZ}, where we show the typical three-fold symmetry of in-plane spin components of thermally occupied states in $k$-space when the precession frequency $\omega$ is set by the Bychkov-Rashba and Dresselhaus fields with $\hbar\omega=\sqrt{(\beta_c k)^2+(\gamma_c k^3\sin 3\phi)^2}$, and the effective spin-orbit field orientation varies as $\tan\theta_B=\beta_c /(\gamma_c k^2 \sin 3\phi)$.

The ensemble average over the out-of-plane component $\langle S_z(t) \rangle$ does {\em{not}} vanish, except for precisely timed gate pulses. Summing over all filled conduction electron states in $\mathbf{k}$-space (again assuming Boltzmann statistics), we have
\begin{align}
%\langle S_z(t)\rangle&=\frac{1}{n}\int_0^{2\pi}\int_0^\infty D_{2d} S_z(\omega t,\theta_B) e^{-\frac{E-E_F}{k_BT}} dE\notag\\
\langle S_z(t)\rangle=\frac{6C}{\pi} \int_0^{\pi/3}\int_0^\infty S_z(\omega t,\theta_B)   e^{-Ck^2} k dkd\phi\label{eq:proj},
\end{align}
where $C=\frac{\hbar^2}{2m^*k_BT}$, and we have exploited the sixfold symmetry of the Dresselhaus field magnitude in the angular integration bound.% of the final expression. 

Notice that Eq.~(\ref{eq:proj}) is independent of the Fermi energy $E_F$. The result of our calculation is therefore independent of the carrier density (which may change upon application of the Bychkov-Rashba field, due to capacitive field-effect from the gate potential), provided the assumption of nondegenerate Boltzmann statistics remains valid. 

\begin{figure}[t!]
\includegraphics[width=8cm]{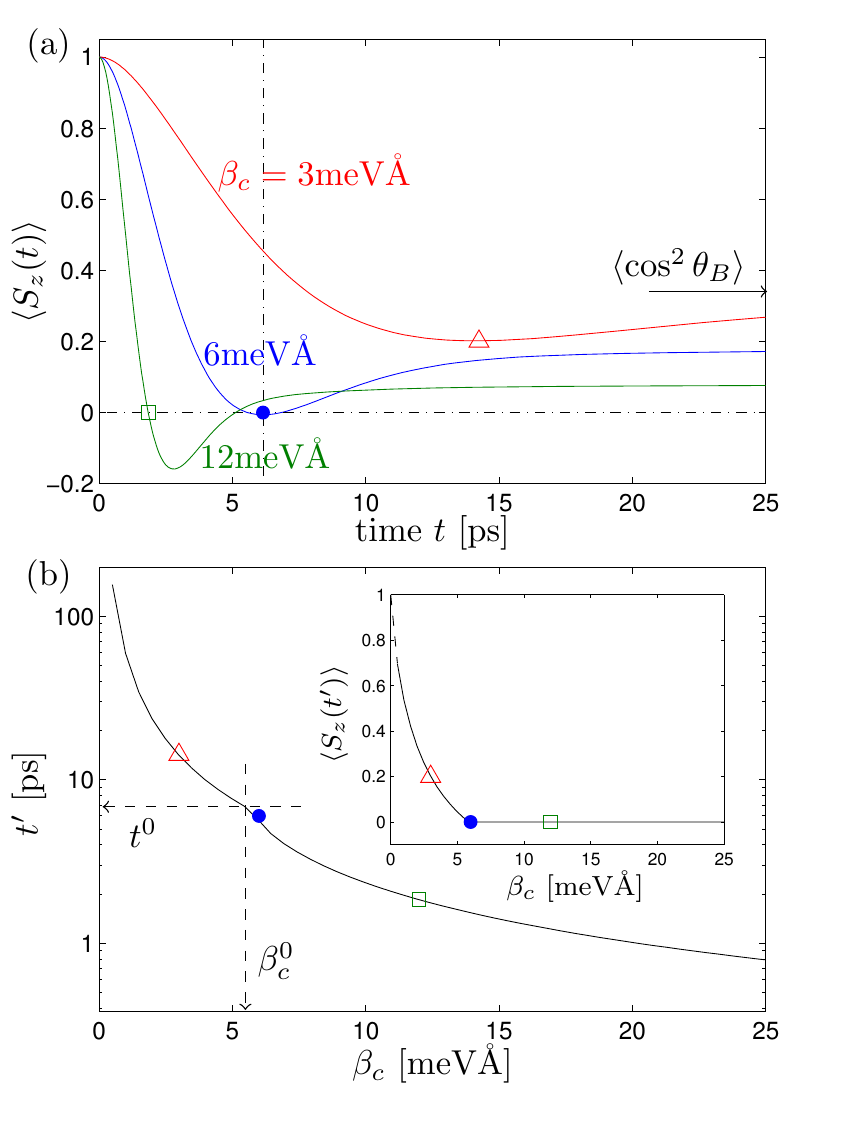}
\vspace{-10pt}
\caption{(a) Time evolution  with $\beta_c=3, 6,$ and $12$~meV\AA. Symbols indicate optimum times (pulsewidths) for minimum out-of-plane projections. (b) Bychkov-Rashba parameter $\beta_c$ dependence of optimized pulsewidth that minimizes this spin projection. Inset: minimum spin projections over the same $\beta_c$ range. Values of $\gamma_c$ and $m^*$ are the same as in Fig. \ref{fig:BZ}.\label{fig:ensemble}}
\end{figure}

Examples of this time evolution at $T=300$~K are shown in Fig. \ref{fig:ensemble}(a), for Bychkov-Rashba parameters $\beta_c=$3, 6, and 12~meV\AA\, (generated by electric fields $\mathcal{E}\approx 1-5$~MV/cm), $\gamma_c=1.044$~eV\AA$^3$ and $m^*=0.655m_0$, as obtained from a tight-binding model for GaSe.\cite{Li_PRB15} The out of plane spin projection $\langle S_z(t)\rangle$ initially decreases, but only for $\beta_c$ larger than a critical value $\beta^0_c\approx6$~meV\AA\, will it vanish completely (at an optimal time $t'<t^0\simeq 7$~ps). This value can be approximated by the condition $B_{BR}=B_D$ for thermal electrons, at $\beta_c^0\simeq\gamma_c\frac{2m^*k_BT}{\hbar^2}$ and $t^0\simeq\pi\hbar^2/\beta_c^0\sqrt{2m^*k_BT}$. If the in-plane Bychkov-Rashba field disappears at the end of an electric field pulse of this duration, the ensemble will remain unpolarized and spin channel reset will be achieved.

Beyond $t>t'$, the spin projections for $\beta_c>\beta_c^0$ undergo a damped oscillation, becoming negative before passing through zero again and saturating at a positive value. The asymptotic values of these spin projections as $t\rightarrow\infty$ correspond to the case where spins are fully dephased, and ensemble averages $\langle\cos\omega t,\sin\omega t \rangle=0$. In other words, only the incoherent part of the spin projections [second term in Eq.~(\ref{eq:sz})] remain.\cite{Li_APL08,*Huang_APL08}

We can calculate the optimum time $t'$ for a range of Bychkov-Rashba parameter $\beta_c$ values as shown in Fig.~\ref{fig:ensemble}(b). For very small values of $\beta_c<\beta_c^0$, when $t'>t^0$, the ensemble out-of-plane spin component never reaches zero. In this case, our calculated $t'$ corresponds to the minimum $\langle S_z\rangle$. Using parameters appropriate for GaSe, this constraint sets a minimum gate-induced electric field of $\approx 1$~MV/cm, consistent with our previous calculation comparing the magnitudes of Dresselhaus and Bychkov-Rashba terms.

\section{discussion}

The short gate pulses of only several picoseconds suggested here set a lower bound for the speed of digital spintronic devices making use of the proposed mechanism. However, this coherent precession scheme assumes that carriers are in the collisionless limit set by the momentum scattering time upper bound. In practice, longer gate pulses (and correspondingly lower perpendicular electric fields) will likely be more practical; if this duration is maintained far longer than the momentum scattering time, a DP-like dephasing and ensemble depolarization will accomplish a similar result. 

However long the gate pulse duration, its rising edge must be abrupt to induce the coherent precession we model. If the gate rise-time is substantially more than the precession frequency, the initially perpendicular spins will simply follow the instantaneous spin-orbit field via adiabatic passage; full depolarization of the out-of-plane spin will then be impossible. 

For this scheme to work, it is essential that there exist occupied regions in $k$-space where the magnitude of Bychkov-Rashba field is greater than Dresselhaus field. This statement does not necessarily imply that materials with the smallest Dresselhaus coefficient should be sought: a moderate value stabilizes spins against dephasing from fluctuating spin-orbit fields generated by e.g. inhomogeneous strain.\cite{Vozmediano_PhysRep10} We thus suggest that, under the right conditions, other platforms with the right configuration of spin-orbit coupling, such as zincblende [110] QWs with a lowest-order Dresselhaus term $\propto k\cos\phi\sigma_z$, may also be effective in enabling spin-channel reset by controlled depolarization.  

%Although we have assumed diffusive transport in thermal equilibrium, the proposed scheme can be implemented even in the nonequilibrium ballistic regime 
%\vspace{1in}
\begin{acknowledgments}
This work was supported by the Office of Naval Research under contract N000141410317, the National Science Foundation under contract ECCS-1231855, and the Defense Threat Reduction Agency under contract HDTRA1-13-1-0013.
\end{acknowledgments}

\bibliography{bib}

\end{document}